\documentclass[a4paper,12pt,twoside,DIV=13,BCOR14mm]{scrartcl}

\usepackage{graphicx,flafter,afterpage,ulem,rotating}
\usepackage[font=small,labelfont=bf,belowskip=2ex]{caption}
\usepackage{textcomp,units}

\usepackage[section]{placeins}

\newcommand{\xH}{\textsuperscript{1}H}

\newcommand{\xP}{\textsuperscript{31}P}
\newcommand{\pp}{\textsuperscript{31}P}

\newcommand{\Tr}{\ensuremath{T_\mathrm{R}}}
\newcommand{\Te}{\ensuremath{T_\mathrm{E}}}
\newcommand{\Tl}{\ensuremath{T_1}}

\newcommand{\taupcr}{\ensuremath{\tau_\mathrm{\,PCr}}}
\newcommand{\Qmax}{\ensuremath{Q_\mathrm{max}}}

\usepackage{color}
\definecolor{medblue}{rgb}{0,0,0.5} \definecolor{medgreen}{rgb}{0,0.5,0}
\definecolor{medred}{rgb}{0.5,0,0} \definecolor{lila}{rgb}{0.3,0,0.3} 
\usepackage{color}
\definecolor{medblue}{rgb}{0,0,0.5} \definecolor{medgreen}{rgb}{0,0.5,0}
\definecolor{medred}{rgb}{0.5,0,0} \definecolor{lila}{rgb}{0.3,0,0.3} 
 \usepackage[breaklinks=true,colorlinks=true,linkcolor=medblue,citecolor=medgreen,urlcolor=medblue]{hyperref}

\newlength{\figwidth}
\newlength{\figborder}
\newlength{\outerfigwidth}
\def\fps@figure{htbp}    

\usepackage{scrpage2}
\lohead{Dynamic \xP\ MRS in soleus muscle}

\usepackage{cite}

\begin{document}
\thispagestyle{empty}
\section*{Localized Semi-LASER Dynamic
    \xP\ Magnetic Resonance Spectroscopy of the Soleus During and
    Following Exercise at 7\,T}

Georg~B.~Fiedler\textsuperscript{$\dagger$ 1,2},
Martin~Meyerspeer\textsuperscript{$\dagger$\,* 1,2}, 
Albrecht~I.~Schmid\textsuperscript{1,2},
Sigrun~Goluch\textsuperscript{1,2},
Kiril~Schewzow\textsuperscript{1,2},
Elmar~Laistler\textsuperscript{1,2},
Arash~Mirzahosseini\textsuperscript{3,4},
Fabian~Niess\textsuperscript{1,2,5},
Ewald~Unger\textsuperscript{1},
Michael~Wolzt\textsuperscript{6},
Ewald~Moser\textsuperscript{1,2}\\[1ex]

\noindent\textsuperscript{$\dagger$}equal contribution\\[3ex]

{\small
\noindent\textsuperscript{1}Center for Medical Physics and Biomedical
Engineering, \\ Medical University of Vienna, Austria\\
\textsuperscript{2}MR Centre of Excellence, Medical
University of Vienna, Austria\\
\textsuperscript{3}Department of Pharmaceutical Chemistry,
Semmelweis University, Budapest, Hungary\\
\textsuperscript{4}Research Group of Drugs of Abuse and Doping Agents, Hungarian
Academy of Sciences, Budapest, Hungary\\
\textsuperscript{5}Graz University of Technology, Institute of Medical
Engineering, Austria\\
\textsuperscript{6}Department of Clinical Pharmacology, Medical
University of Vienna, Austria\\
}

\noindent
DOI: \href{http://dx.doi.org/10.1007/s10334-015-0484-5}{10.1007/s10334-015-0484-5}\\
Original version on publisher's website:\\ \url{http://link.springer.com/article/10.1007/s10334-015-0484-5}\\


\paragraph{\textsuperscript{*}Correspondence to:}\quad\\
Martin~Meyerspeer, PhD\\
\href{mailto:martin.meyerspeer@meduniwien.ac.at}{martin.meyerspeer@meduniwien.ac.at}\\
Center for Medical Physics and Biomedical Engineering\\
Medical University of Vienna\\
Lazarettgasse 14\\
1090 Wien, Austria\\
+43 (0)1 / 40400 - 64610

\newpage


\subsection*{Abstract}

\subsubsection*{Object}

This study demonstrates the applicability of semi-LASER localized
dynamic \xP\,MRS to deeper lying areas of the exercising human soleus
muscle (SOL). The effect of accurate localization and high temporal
resolution on data specificity is investigated.

\subsubsection*{Materials and Methods} 

To achieve high signal-to-noise ratio (SNR) at a
temporal resolution of 6\,s, a custom-built calf coil array was used at
7T. 
The kinetics of phosphocreatine (PCr) 
and intracellular pH were quantified separately in SOL and
gastrocnemius medialis (GM) muscle of 9 volunteers, during rest,
plantar flexion exercise and recovery.

\subsubsection*{Results} 

The average SNR of PCr at rest was 64$\pm$15 in SOL (83$\pm$12 in GM).
End exercise PCr depletion in SOL (19$\pm$9\%) was far lower than in
GM (74$\pm$14\%).  pH in SOL increased rapidly
and, in contrast to GM, remained elevated until the end of exercise.

\subsubsection*{Conclusion} 

\xP\,MRS in single-shots every 6\,s localized in the deeper lying SOL
enabled quantification of PCr recovery
times at low depletions and of fast pH changes, like the initial
rise.  
Both high temporal resolution and accurate spatial localization improve
specificity of Pi and thus pH quantification by avoiding multiple, and
potentially indistinguishable sources for changing Pi peak shape.

\vfill
{\small
  \paragraph{Keywords:} Skeletal Muscle, Phosphocreatine,
	In Vivo NMR Spectroscopy, Soleus Muscle, Physical Exertion
	}

\clearpage

\section*{Introduction}

Phosphorus-31 nuclear magnetic resonance spectroscopy (\pp\,MRS) is a
well-established, powerful tool for noninvasive measurements of energy
metabolism in exercising muscle \cite{Hoult:1974, Chance:2006,
Bendahan:2004, Prompers:2006, Kemp-Meyerspeer:2007}. Particularly, the
concentrations and kinetics of phosphorylated creatine (PCr),
inorganic phosphate (Pi) and the intracellular pH \cite{Moon:1973},
are accessible, which are directly connected to ATP turnover via
the creatine kinase reactions. 
Numerous dynamic studies have been conducted in the human calf muscle
using plantar flexion exercise \cite{Prompers:2006,Chance:2006}, where
the main contributing muscles are gastrocnemius medialis (GM),
gastrocnemius lateralis (GL) and soleus (SOL), all of which act on the
Achilles tendon. GM and GL are located superficially and originate
from the lower end of the femur \cite{Gray:1995}. The deeper lying SOL
is connected to tibia and fibula \cite{Gray:1995}, and slightly
differs in function. Hence its activation during plantar flexion is
also different from the gastrocnemius muscles, the proportion of
recruitment depending, amongst others, on the angle of the knee
\cite{Price:2003}.
Based on the differences of these muscles regarding function and
anatomy, PCr and Pi kinetics are also expected to differ
\cite{Vandenborne:1993, Noseworthy:2003, Vandenborne:2000,
Jacobi:2012, Layec:2013, Forbes:2009c}.  Acquiring nonlocalized data from a
mixture of muscle tissues implies averaging across their respective
metabolite concentrations, exchange rates and pH. 
This degrades sensitivity to small metabolic changes, especially when
averaging across muscles with different levels of activation, and does
not allow assessing specific metabolic properties from muscles with
heterogeneous phenotype (e.g. plantar flexor muscles).
This has been confirmed recently, comparing standard nonlocalized MRS
of the calf with a volume of interest localized in GM, yielding
different results \cite{Meyerspeer:2012}.
Within each muscle, on the other hand, activation is relatively
homogeneous \cite{Davis_ismrm:2013}, allowing single-voxel spectroscopy
to be applied using a large, anatomy-matched volume.

Several localization schemes for  \xP\ MRS have been used in muscle
metabolism studies, the simplest using the limited sensitive volume of
a small coil, 
further methods include ISIS \cite{Vandenborne:1993, Allen:1997} DRESS
\cite{Valkovic:2014}, and STEAM \cite{Meyerspeer:2005}.  
Small coil localization and DRESS cannot monitor deeper lying
muscles by virtue, and STEAM inherently loses half the signal
\cite{Meyerspeer:2011}. 
ISIS needs at least eight scans for 3D localization, and
with a \Tl\ of PCr of approximately 4\,s at 7\,T \cite{Bogner:2009},
this results in low temporal resolution.
Methods like spectrally selective \pp\,MRI or chemical shift imaging
(CSI) are rapidly advancing. With these methods, several muscles can
be measured simultaneously and potential metabolic heterogeneities
within a muscle can be detected, however, within their limits of
point-spread function, spatial resolution, SNR and temporal
resolution.
Parasoglou et al. \cite{Parasoglou:2012a} showed the feasibility of 3D
\xP\ MRI to quantify PCr dynamics in the human calf. One serious
drawback of this method is that localized pH is not accessible by PCr
imaging,
which is an important parameter for characterization of aerobic and
glycolytic muscle metabolism. 
CSI, on the other hand, suffers from poor voxel definition and
requires far longer acquisition times. While metabolic processes on
the order of 20 -- 30 seconds have been observed using a gated CSI
protocol \cite{Forbes:2009c},  this approach depends on reproducibly
reaching a steady-state (via a known dependence on time, e.g.,
mono-exponential) and imposes restrictions, e.g., acidosis
should not be induced.
The semi-LASER single-voxel MRS acquisition scheme is applicable to
any exercise intensity and metabolic state, also any orientation and
size of an individual muscle can be covered by the free
positioning of the double oblique voxel in three dimensions, provided
sufficient $B_1$.  Further, the contamination by signal from outside
of the VOI with the semi-LASER
sequence is very low at ca. 1\% \cite{Meyerspeer:2011}.

The location of SOL, which is covered by the neighboring gastrocnemius
muscles, renders data acquisition more challenging. 
The aim of this study was to acquire accurately localized \xP\ MR
spectra from exercising soleus and to show that it is possible to
follow Pi and thus intracellular pH throughout rest, exercise and
recovery with a temporal resolution of 6\,s, as well as to quantify
the PCr recovery time in this muscle via a monoexponential fit.

To our knowledge, this has so far not
been demonstrated for \xP\,MRS localized to a deeper muscle area.
To achieve this goal,
the combined SNR advances from an ultra-high field scanner, a dedicated,
multi-channel calf-coil and an adapted semi-LASER localization scheme
were exploited.
This extends previous studies (e.g.\,\cite{Meyerspeer:2012}) by
demonstrating that soleus muscle can be monitored under exercise with
adequate data quality, comparable to previous results from
gastrocnemius.
%


\section*{Materials and Methods}

\begin{figure}[\figwidth]
  \centering
  \includegraphics[height=0.28\textheight]{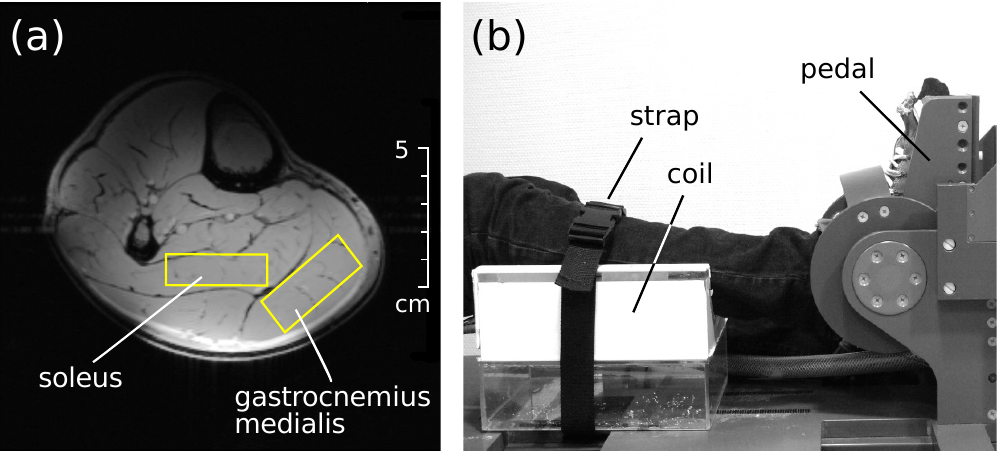}
  \caption{(a) Typical position of the voxels in soleus and
    gastrocnemius medialis muscle, respectively, for localized \xP\
    MRS performed during two consecutive plantar flexion exercise
    bouts.  (b) Coil position and shape, together with a fixation
    strap, helped to keep the calf in shape even during exercise with
    the ergometer.  }
  \label{fig:voxelplacement}
\end{figure}

Eleven young healthy subjects (age: $26 \pm 5$ years, body mass index:
$22 \pm 1$\,kg/m$^2$, $4$ females) were scanned after written informed
consent, according to the guidelines of the local ethics committee and
the declaration of Helsinki. Nine subjects completed the full protocol
with data acquisition in SOL and GM, two subjects were measured for
assessment of reproducibility at low PCr depletions in SOL only.
An in-house built \xH /\xP\ transceive coil array was used in a
\unit[7]{T} whole body MR scanner (Siemens Medical, Erlangen,
Germany). The coil consisted of two channels for \xH\ and three
channels for \xP\, and was shaped to a half cylinder ($d =
\unit[14]{cm}$, $l = \unit[10]{cm}$) to accommodate a human calf
\cite{Goluch:2014}.
The semi-LASER sequence \cite{Meyerspeer:2011,Scheenen:2008b} was
applied with a repetition time (\Tr ) of 6\,s.  As semi-LASER is a
single-shot sequence, this was also the true temporal resolution of
the experiment.  The shim and the transmit voltage of the RF pulses
were adjusted individually for each subject and voxel position, as
described previously \cite{Meyerspeer:2011}.
The acquisition of spectra from the deeper-lying SOL requires more
transmit power (at a given pulse duration) to achieve full adiabatic
inversion of the spins, compared to the better studied superficial GM.
Limited by the available RF transmit power, the inversion pulse
duration was typically 4600\,\textmu s in SOL (3400\,\textmu s for
voxels located in GM).  The shortest possible echo time (\Te) was
selected, which was typically 29\,ms for SOL 
(24\,ms for GM). 
To optimize SNR while ensuring specificity to single muscles, volumes
of interest (VOI) were positioned double obliquely and chosen as large
as possible, while fitting entirely within the respective muscle.
This was done individually for each subject, using 25 transversal
gradient echo MRI slices for voxel planning.
The resulting average volumes were VOI$_\mathrm{SOL}$ =
{45\,$\pm$\,9\,cm\textsuperscript{3} {\normalsize
    (66\,$\times$\,44\,$\times$\,16\,mm\textsuperscript{3})}} and
VOI$_\mathrm{GM}$ = {40\,$\pm$\,5\,cm\textsuperscript{3} {\normalsize
(63\,$\times$\,41\,$\times$\,16\,mm\textsuperscript{3})}.}
Typical voxel positions can be seen in Fig.~\ref{fig:voxelplacement}a.
Acquired spectra were processed with the jMRUI software package
\cite{Naressi:2001}, using the AMARES time domain fit algorithm
\cite{Vanhamme:1997}. %
PCr and Pi signal intensities, representing the area under the peak in
the frequency domain, were normalized to baseline PCr concentrations.
pH values were calculated from the chemical shift between Pi and PCr
\cite{Moon:1973}, and PCr recovery times \taupcr\ were fitted
mono-exponentially to the PCr recovery time course.  Maximum oxidative
capacity (\Qmax) was calculated from \taupcr\ and the end-exercise
values PCr and pH, using the ADP driven model
\cite{Kemp:1994,Schmid:2014}, errors of \Qmax\ were calculated by
error propagation.
SNR was quantified as the ratio of the PCr peak amplitude to the
standard deviation (SD) of noise at around 10 ppm up-field from the
PCr frequency (width of noise region: 1/8 of total bandwidth).

The measurement protocol comprised 2~minutes at rest, during which
time baseline data were acquired, 5~minutes of 
submaximal plantar flexion, and 7~minutes of recovery time.
The isotonic exercise on a custom built MR compatible ergometer
consisted of two consecutive pedal pushes between each period of MR
data acquisition, which was every $\Tr = \unit[6]{seconds}$, i.e., the
effective frequency of the exercise was \unit[0.3]{Hz}.
Maximum voluntary contraction force (MVC) of each subject was measured in
the ergometer on the scanner bed, prior to the MRS measurements. 
The force of the exercises was set to 40\,\% of MVC, via the pressure in the
air piston of the ergometer, and controlled throughout the experiment. 
The knee was straight and the calf was placed on the form-fitted array
coil and fixed in position using a strap across the tibia (see
Fig.~\ref{fig:voxelplacement}b).
Two equivalent exercise bouts were carried out, the first for
measuring the time course of localized \pp\ MR spectra in GM, and the
second for SOL. They were separated by 28 minutes of inactivity (on
average, minimum was 20 minutes), ensuring that PCr, Pi and pH values
had fully recovered to basal levels at the beginning of the second
bout.

Control experiments were performed in four subjects (two from the
group performing the full protocol, and two additional subjects) by
repeatedly measuring in SOL. The purpose of these measurements, was to
assess the reproducibility of metabolite quantification and fitting of
\taupcr\ at low PCr depletion in this muscle between subsequent
exercise bouts, with a resting phase of 25\,min.  Day-to-day
reproducibility was tested after one week in one subject.


\section*{Results}

PCr and Pi were quantified from \xP\ spectra acquired in time series
with a temporal resolution of 6\,s.
The average SNR 
of PCr at rest
was $64\pm15$ in SOL, and $83\pm12$ in GM.
The mean force applied by the volunteers during exercise, as measured
with the sensor on the ergometer pedal, was {$40 \pm 10$\,\% of MVC in the SOL
bout, and $40 \pm 8$\,\% of MVC in the GM bout.}  The mismatch between
each individual's forces in the two exercise bouts was {$1.8 \pm
1.3$\,\%} of MVC. All values relate to individual MVC.

\begin{figure}
  \begin{center}
    \vspace{1ex}
    \includegraphics[width=4.5in]{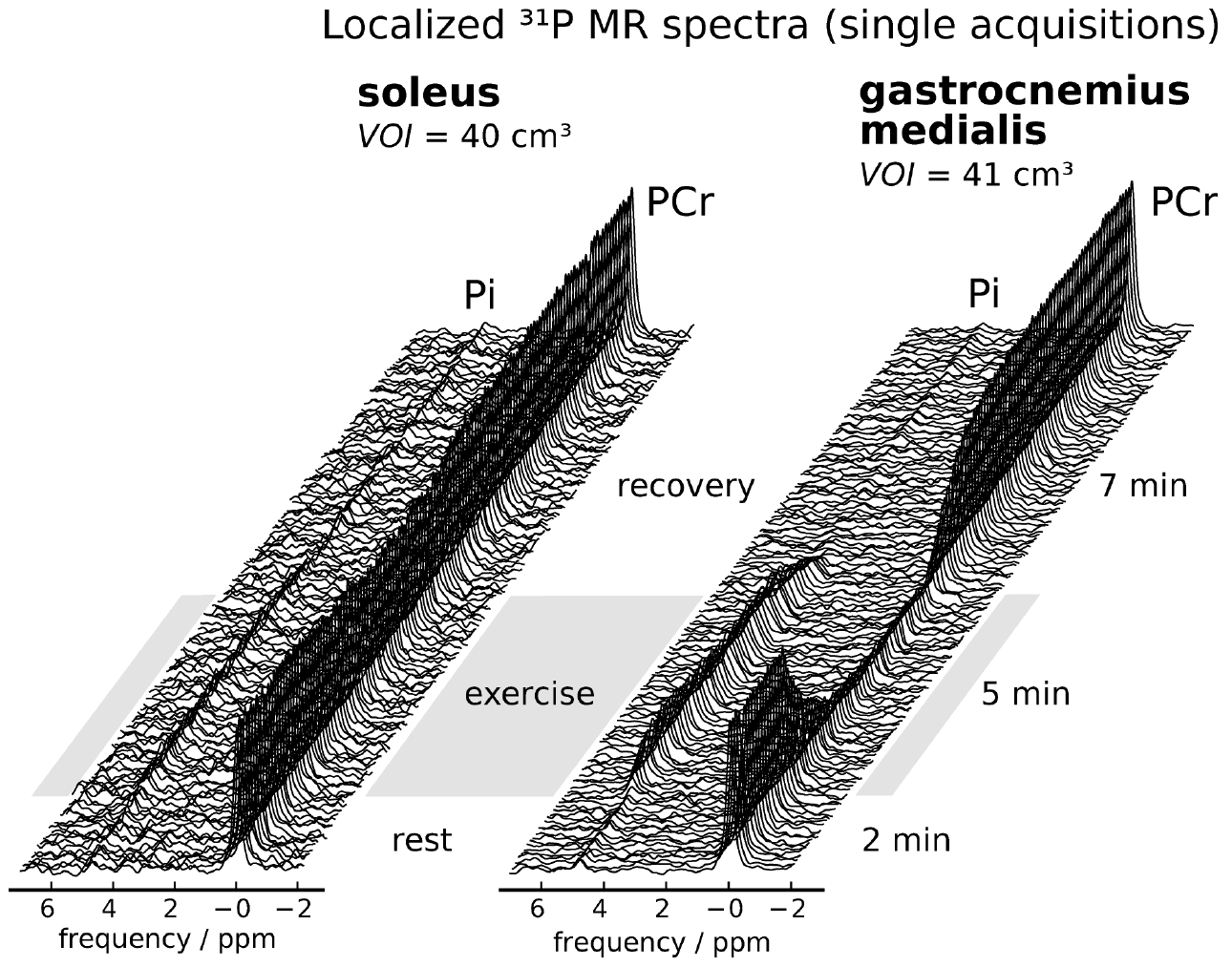}
  \end{center}
  \caption{Stack plots of spectra acquired in SOL (left) and GM
    (right) during rest, plantar flexion exercise (gray area) and
    recovery of one subject. 
    Single-shot spectra ($\Tr = \unit[6]{s}$) are shown without
    averaging. Exponential line broadening of 25\,Hz and zero filling
    ($8\times$) were applied for better visualization.
    \label{fig:stackplots}}
\end{figure}

\begin{figure}[\figwidth]
  \begin{center}
    \includegraphics[width=0.8\textwidth]{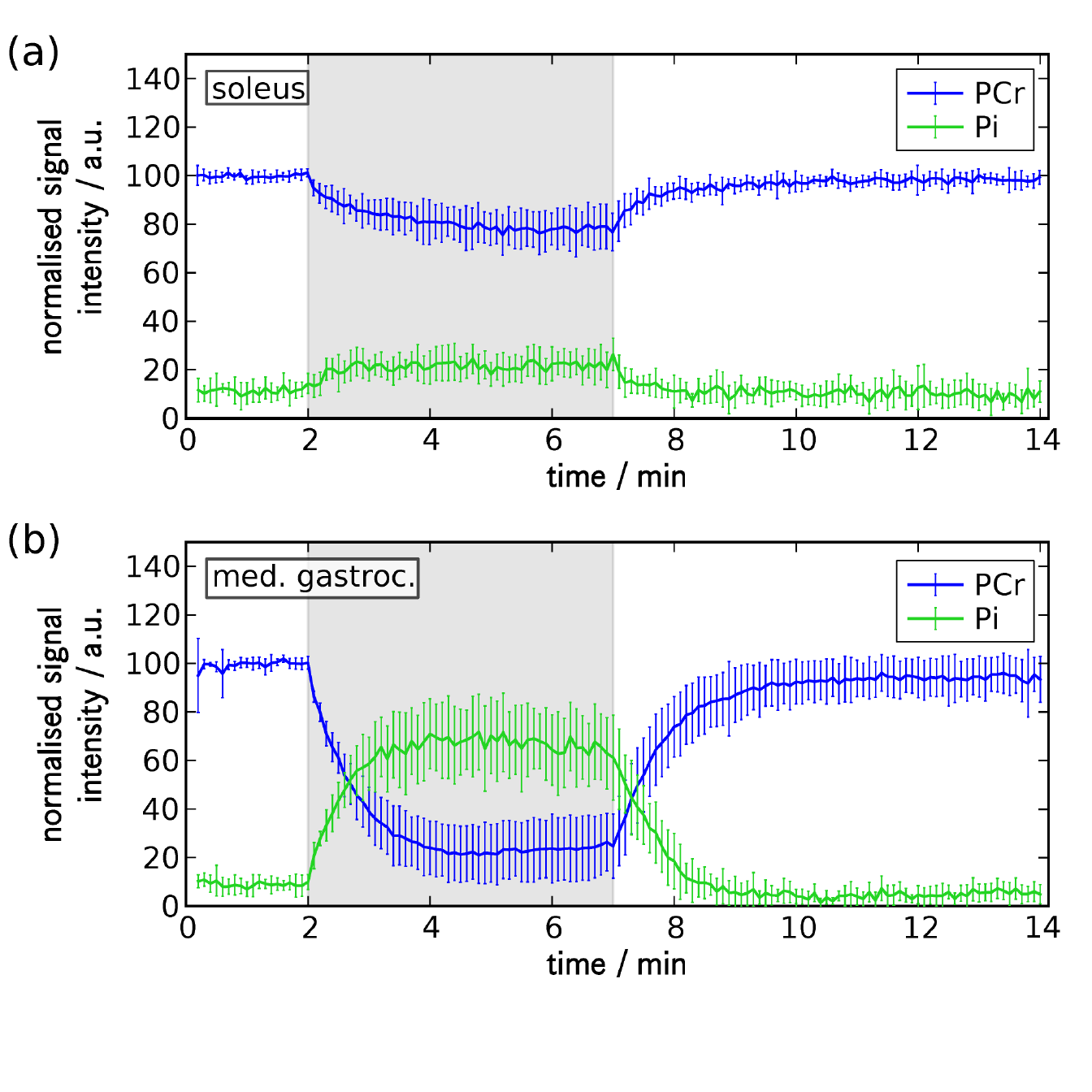}
    \caption{PCr and Pi signal in (a) SOL and, for comparison, (b) GM
      at rest, during exercise (marked Gray) and during recovery, from
      two equivalent exercise bouts. Mean $\pm$ SD of $n=9$
      subjects.}
    \label{fig:pcr-sol-all-sum} 
  \end{center} 
\end{figure}

Stack plots of \pp-spectra from a single subject are shown in
Fig.~\ref{fig:stackplots}. Data are plotted without temporal averaging,
representing the acquisition's full time resolution of~6\,s.
Time courses of PCr and Pi signal averaged over all 9 subjects are
shown in Fig.~\ref{fig:pcr-sol-all-sum}.  PCr depletion was detected
in the soleus muscle of all subjects, and ranged from 6\,\% to 38\,\%
at end of exercise.
GM showed far stronger depletion in all subjects, ranging between 
{49\,\%} and 94\,\%, which is consistent with \cite{Schmid:2014}.
Average end-exercise PCr depletions of SOL and GM can be read from
Fig.~\ref{fig:pcr-sol-all-sum} and are summarized in
Table~\ref{tab:tabgastsol}. The table also shows the average PCr
recovery times \taupcr, excluding the one subject with only 6\,\%
depletion in SOL, due to insufficient dynamic range for the PCr
recovery fit.

\newcommand{\mcc}[1]{\multicolumn{2}{c}{#1}}

\begin{table}
  \begin{center}
    \caption{PCr depletion (relative to resting PCr signal), PCr
      recovery time constant \taupcr, pH values at rest, at end of
      exercise and at the post exercise minimum and maximum oxidative
      capacity \Qmax\ of $n=9$ subjects.  Data were quantified from
      unaveraged spectra, errors for PCr depletion and \taupcr\ of
      individuals are SDs resulting from the fitting algorithm, the
      uncertainty of \Qmax\ was derived via error propagation.
      The pH at the end of exercise and the post exercise
      minimum in GM were quantified from four unaveraged spectra,
      (SDs over five data points). The resting pH in SOL and GM and
      the pH during the last 2\,min measured in SOL was derived from
      blocks of five accumulated spectra (SDs over four data points).
      Group averages represent the mean $\pm$ SD of individuals' data.
    }
    \label{tab:tabgastsol}
		\begin{tabular}{c@{\,}c@{\,}
			r@{$\ \pm\ $}lr@{$\ \pm\ $}lr@{$\ \pm\ $}l
			r@{$\ \pm\ $}lr@{$\ \pm\ $}l@{\hspace*{-4em}}r@{$\ \pm\ $}l}
      & Subj. & \multicolumn{2}{c}{PCr} &
			\mcc{$\tau_\mathrm{\,PCr}$ } 
			&\multicolumn{2}{c}{pH$_\mathrm{rest}$}
			&\multicolumn{2}{c}{pH$_\mathrm{end\,ex}$}
			&\multicolumn{2}{c}{{pH$_\mathrm{end\,rcv}$(SOL)}} 
			&\multicolumn{2}{c}{\hspace*{-1em}{Q$_\mathrm{max}$}} \\ 
      &       & \multicolumn{2}{c}{depletion} &
			\mcc{} 
			&\multicolumn{2}{c}{}
			&\multicolumn{2}{c}{}
			&\multicolumn{2}{c}{pH$_\mathrm{min}$(GM)} 
			&\multicolumn{2}{c}{{\hspace*{-0.5em}\footnotesize[mM/min]}} \\ \hline
\rule[0ex]{0ex}{2.5ex}
SOL     &1	&6&1\,\%	&\mcc{-}	&7.04&0.004	&7.05&0.01	&7.03&0.006	&\mcc{-}\\
	&2	&38&2\,\%	&54&5\,s	&6.99&0.014 	&7.03&0.01	&7.00&0.010	&23&2\\
	&3	&23&2\,\%	&23&4\,s	&7.03&0.006 	&7.02&0.02	&6.99&0.005	&40&7\\
	&4	&17&1\,\%	&20&3\,s	&7.00&0.002	&7.02&0.01	&6.97&0.014	&39&7\\
	&5	&14&1\,\%	&63&12\,s	&7.02&0.001	&7.05&0.01	&7.00&0.002	&10&2\\
	&6	&20&3\,\%	&24&7\,s	&7.02&0.003	&7.08&0.01	&7.01&0.002	&33&10\\
	&7	&15&2\,\%	&30&9\,s	&7.02&0.010	&7.07&0.01	&7.00&0.018	&23&7\\
	&8	&11&1\,\%	&77&15\,s	&7.02&0.006	&7.06&0.02	&7.00&0.012	&8&2\\
	&9	&26&1\,\%	&42&5\,s	&7.02&0.004	&7.06&0.01	&7.01&0.006	&23&2\\
\hline\rule[0ex]{0ex}{2.5ex}
GM      &1	&65&1\,\%	&32&1\,s 	&7.03&0.003	&7.03&0.01	&6.90&0.05	&51&1\\
	&2	&90&2\,\%	&71&3\,s	&6.98&0.019	&6.47&0.03	&6.38&0.06	&30&1\\
	&3	&94&1\,\%	&68&1\,s	&7.08&0.003	&6.58&0.02	&6.46&0.03	&30&1\\
	&4	&69&1\,\%	&34&1\,s	&7.01&0.002	&6.97&0.02	&6.80&0.01	&49&1\\
	&5	&69&1\,\%	&62&2\,s	&7.05&0.010 	&6.78&0.01	&6.65&0.10	&29&1\\
	&6	&76&2\,\%	&44&2\,s	&7.04&0.005	&6.95&0.03	&6.83&0.12	&41&2\\
	&7	&49&1\,\%	&45&1\,s	&7.04&0.004	&7.01&0.01	&6.87&0.02	&31&1\\
	&8	&83&1\,\%	&45&1\,s	&7.05&0.008	&6.87&0.01	&6.68&0.01	&42&1\\
	&9	&73&2\,\%	&37&2\,s	&7.06&0.007	&6.93&0.02	&6.75&0.03	&48&1\\
\hline\rule[0ex]{0ex}{2.5ex}
SOL	&	&19&9\,\%	&37&17\,s	&7.02&0.02	&7.06&0.02	&7.00&0.01	&25&12\\
GM	&	&74&14\,\%	&49&15\,s	&7.04&0.03	&6.85&0.19	&6.70&0.19	&39&9\\
   \end{tabular}
  \end{center} 
\end{table}

\begin{figure}
  \begin{center}
    \includegraphics[width=0.7\textwidth]{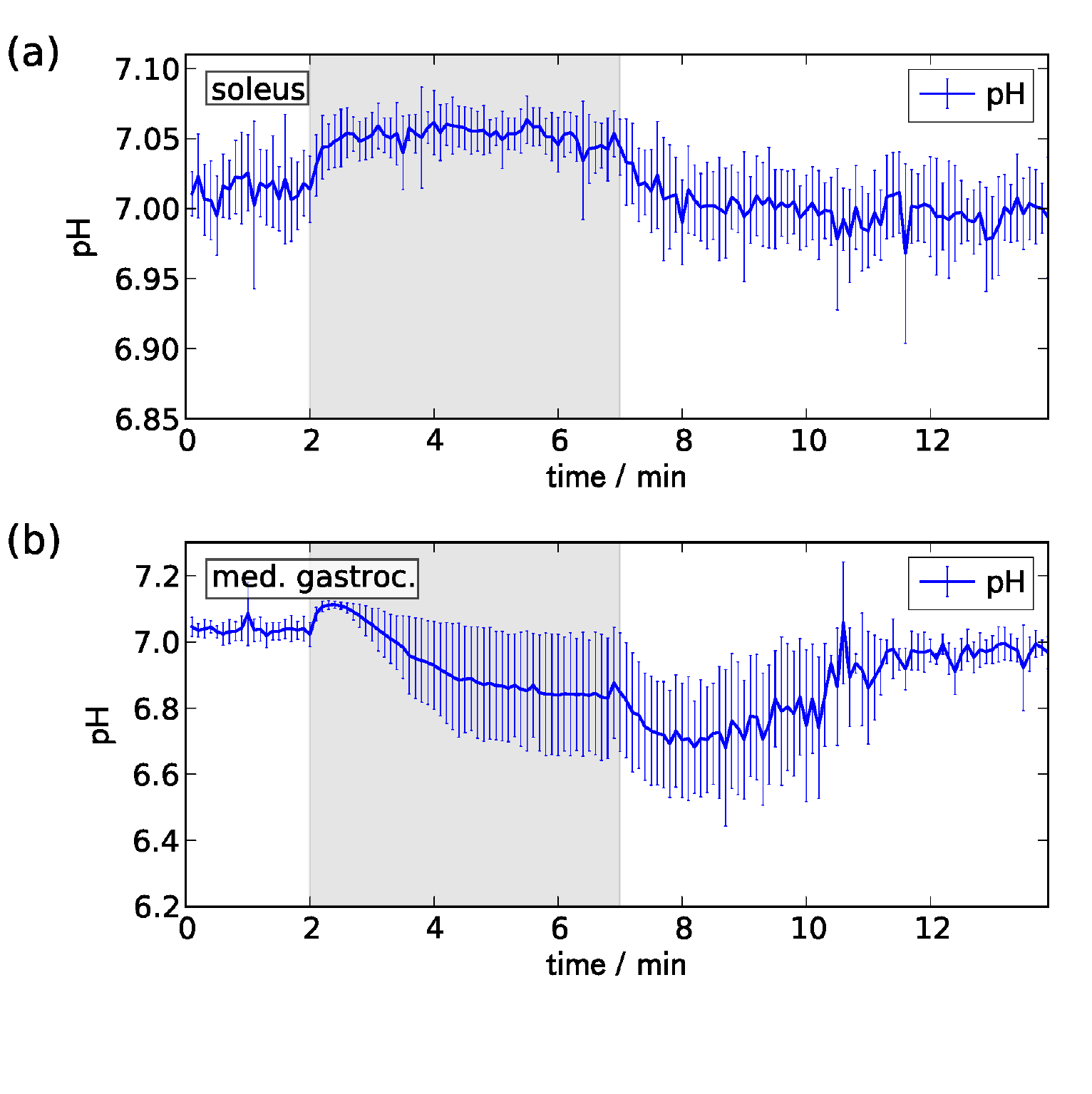}
    \caption{Intracellular pH in (a) SOL and (b) GM with a time resolution
      of 6\,s, mean $\pm$ SD of all $n = 9$ subjects.  Note the different
      scale of the two plots. }
    \label{fig:pH} 
  \end{center} 
\end{figure}

Intracellular pH was also quantified at a temporal resolution of 6\,s.
The time courses of pH in SOL (Fig.~\ref{fig:pH}a) and GM
(Fig.~\ref{fig:pH}b), respectively, are shown as the average over all
9 subjects. Data points where the SNR of Pi was below 1.3 for
individual subjects were not included.  Average pH values at rest, end
of exercise and the post-exercise minimum in SOL and GM are given in
Table~\ref{tab:tabgastsol}.

In SOL, pH increased immediately at the onset of exercise, rising by
$0.03 \pm 0.01$ within 12\,s, 
see Fig.~\ref{fig:pH}a. The pH remained elevated throughout the
exercise, with a maximum of $7.09 \pm 0.02$. 
During recovery, pH returned to resting values within {$23
\pm 8$\,s
and continued to decrease slightly below resting values by
$-0.016\pm0.012$ ($p < 0.005$ pre vs.\ post exercise pH), as
quantified from 4 blocks of 5 averaged spectra at rest and during the
last 2 minutes of the measurement.
The pH evolution in SOL} was in stark contrast to GM:
the initial pH rise in the first 12\,s was $0.07 \pm 0.02$ (see
Fig.~\ref{fig:pH}b), and the maximum of $7.12 \pm 0.01$ was reached
after $23 \pm 8$\,s. 
End-exercise pH in GM varied widely between subjects, ranging from
6.47 to 7.02 and was followed by a post-exercise drop of
$0.15 \pm 0.03$, before returning towards its resting value.

\begin{figure}[\figwidth]
  \begin{center}
    \includegraphics[width=1.0\textwidth]{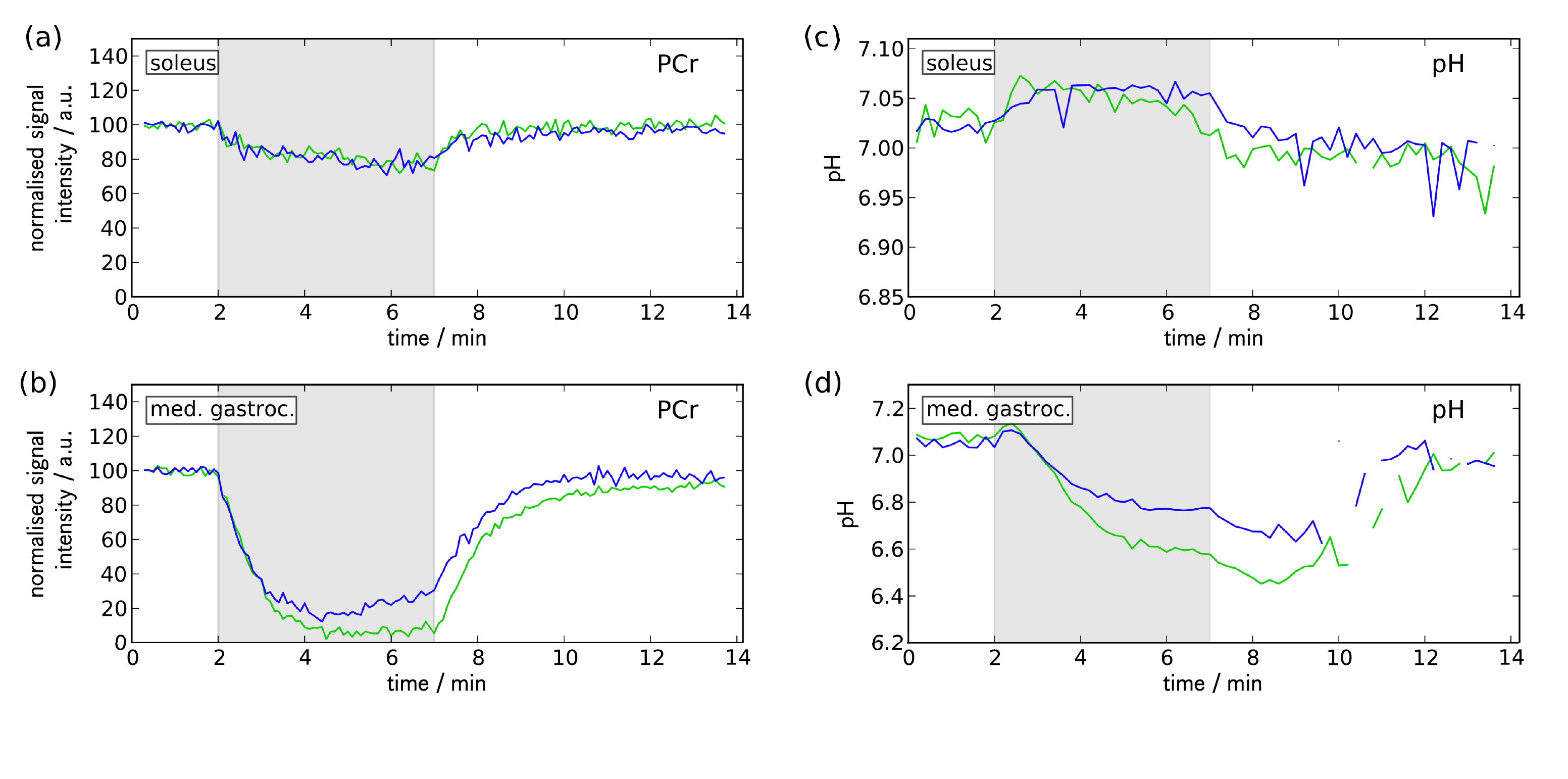}
    \caption{Time courses of PCr signal 
      and pH evolution of the same two representative subjects.
      The PCr signal in (a) SOL and (b) GM is shown with 6\,s time
      resolution. The pH evolution shown in (c) for SOL and (d) for GM
      was derived from spectra after averaging two time points each
      (i.e.,\ 12\,s time resolution), datapoints with an SNR of the Pi
      peak below 1.3 are omitted.  Note that the plots showing pH (c,
      d) are scaled differently.}
    \label{fig:indiv-depl} 
  \end{center} 
\end{figure}
Two representative individual subjects' datasets are presented in
Fig.~\ref{fig:indiv-depl}, showing PCr time courses
(Fig.~\ref{fig:indiv-depl}a and b) with a temporal resolution of 6\,s.
For better visualization only, the pH is shown in
Fig.~\ref{fig:indiv-depl}c and d with 12\,s time resolution, obtained
by averaging two successive spectra at each data point. 
As for averaged data, points with $\mathrm{SNR(Pi)} < 1.3$ were 
omitted in the individual pH graph.

\begin{figure}[\figwidth]
  \begin{center}
    \includegraphics[width=0.9\textwidth]{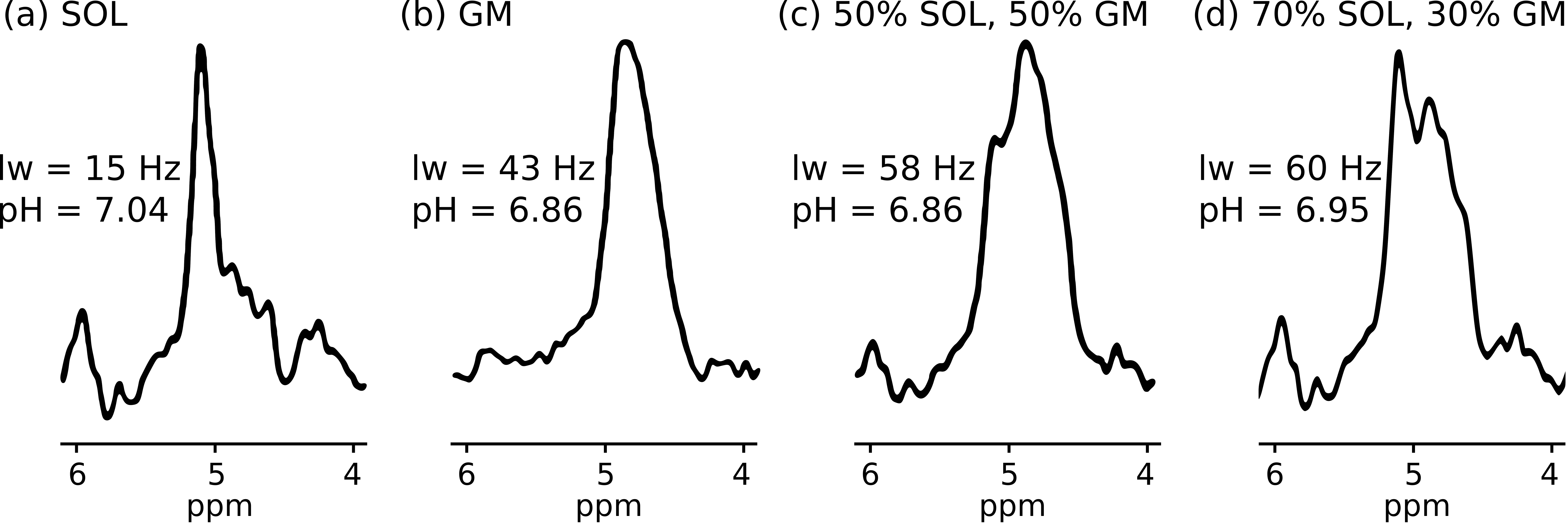}
    \caption{Pi peaks in spectra from (a) SOL, (b) GM, and averages of
      the two datasets, (c) 50\,\% SOL and 50\,\% GM, (d) 70\,\% SOL
      and 30\,\% GM.  All spectra were measured in the same subject,
      represent the same point in time (after 1 minute of exercise),
      and are scaled to the same amplitude.  Linewidth (lw) and the
      calculated pH are given.}
    \label{fig:var-avg}
  \end{center} 
\end{figure}

To investigate possible effects on the measured Pi peak by low spatial
resolution or a partial volume effect, retrospective spatial averages
of spectra are compared to the original data. Fig.~\ref{fig:var-avg}
(a) and (b) show spectra localized in SOL and GM, respectively, from
unaveraged single-shot acquisitions, after 1 minute of exercise.  To
simulate partial volume effects, Fig.~\ref{fig:var-avg} (c) shows the
average of the spectra from the two muscles, while
Fig.~\ref{fig:var-avg} (d) shows a combined spectrum obtained by
weighting SOL and GM data by 70:30.  All data in
Fig.~\ref{fig:var-avg} are from the same subject.  Depending on the
effective time interval of temporal averaging, the width and shape of
the Pi peak also varies, corresponding to different derived pH (data
not shown).

Results of the measurements repeated in the same subjects show good
reproducibility of the applied force, which was adjusted to yield
relatively low PCr depletion in SOL of 22\,\% and 26\,\% (group
averages).
Also the end-exercise pH in the neutral to slightly alkaline regime
and PCr recovery time of \unit[20]{s} vs.\ \unit[27]{s} were
reproducible (see Table~\ref{tab:repro}) and not significantly
different between bouts.  PCr depletion and pH changes were also found
to be reproducible in measurements repeated after one week.

\begin{table}
  \begin{center}
    \caption{Repeated measurements in SOL, for each subject with the
      same exercise intensity, respectively. Bouts one and two were
      spaced by 25\,min of inactivity.  The measurements on subject
      \#10 were repeated after one week.}
    \label{tab:repro}
    \begin{tabular}{ccr@{$\ \pm\ $}l r@{$\ \pm\ $}l r@{$\ \pm\ $}l r@{$\ \pm\ $}l
                                     r@{$\ \pm\ $}l r@{$\ \pm\ $}l r@{$\ \pm\ $}l}
      Subject  &Bout
		&\multicolumn{2}{c}{PCr~depl.} 
		&\multicolumn{2}{c}{\taupcr}
		&\multicolumn{2}{c}{pH$_\mathrm{end}$} 
		&\multicolumn{2}{c}{Force\,/\,N}\\\hline                           
\#6\rule[0ex]{0ex}{2.5ex}
		&1	&48&2\,\%	&27&2\,s	&6.92&0.03	& 243&14 \\
		&2	&34&2\,\%	&34&3\,s	&6.98&0.02	& 234&15 \\
\#9\rule[0ex]{0ex}{3.0ex}
		&1	&18&2\,\%	&17&4\,s	&7.04&0.01	& 255&22 \\
		&2	&12&2\,\%	&19&7\,s	&7.04&0.01	& 222&16 \\
\#10\rule[0ex]{0ex}{3.0ex}
		&1	&3&2\,\%	&\mcc{-}	&7.07&0.01	& 291&11 \\
		&2	&4&2\,\%	&\mcc{-}	&7.07&0.02	& 292&10 \\
\#11\rule[0ex]{0ex}{3.0ex}
		&1	&36&1\,\%	&15&2\,s	&7.06&0.01	& 285&24 \\
		&2	&39&2\,\%	&28&3\,s	&7.06&0.01	& 283&29 \\
\hline             
Mean\rule[0ex]{0ex}{2.5ex}
		&1	&26&20\,\%	&20&6\,s	&7.02&0.07	& 267&21 \\
		&2	&22&17\,\%	&27&7\,s	&7.04&0.04	& 256&32 \\
\\
\multicolumn{5}{l}{Measurement repeated after 1 week} \\	[0.07in]	
Subject  &Bout
		&\multicolumn{2}{c}{PCr~depl.} 
		&\multicolumn{2}{c}{\taupcr}
		&\multicolumn{2}{c}{pH$_\mathrm{end}$} 
		&\multicolumn{2}{c}{Force\,/\,N}\\\hline   	
\#10\rule[0ex]{0ex}{2.5ex}
		&1	&6&2\,\%	&\mcc{-}	&7.10&0.02	& 284&9 \\
		&2	&4&4\,\%	&\mcc{-}	&7.11&0.03	& 283&11 \\
	\end{tabular}\\[2ex]
 \end{center} 
\end{table}

These measurements also showed a small, significant pH decrease by
$-0.03\pm0.02$, post vs.\ pre exercise, in all bouts.  The pH measured
during the 2 minutes before exercise was not significantly different
between bout 1 and 2.


\section*{Discussion}

This study successfully shows that quantification of PCr, Pi and pH is
feasible with high specificity, even in the deeper lying human soleus
muscle, by combining high temporal resolution of 6\,s with spatial
selectivity. This was achieved by exploiting the high SNR of \xP\
spectra acquired with a semi-LASER sequence, using a custom-built
form-fitted array coil \cite{Goluch:2014} on a 7T MR scanner, in
combination with a plantar flexion paradigm. 

SOL consistently showed PCr depletion, although at far lower levels
than GM. The time courses of pH differed even qualitatively: sharing a
similar rise at the beginning of exercise, pH in SOL remained alkaline
during the exercise and dropped towards basal levels shortly after the
end of exercise. In GM however, pH was declining throughout the
exercise, with a further drop post-exercise.  This is consistent with
ASL data from Schewzow et al.  \cite{Schewzow:2014}, where perfusion
in GM increased much more than in SOL in a comparable exercise
paradigm.
In a previous study \cite{Meyerspeer:2012}, simple
acquisition of FID vs.\ a localization scheme was compared, 
using the same standard
single loop coil in both measurements, demonstrating significantly
different results from the two methods.
The results from the present study indicate that contamination from
the lesser activated SOL is a possible source of {bias} for
quantification of PCr, Pi and consequently pH in unlocalized
experiments.
This emphasizes the importance of spatial localization in order to
provide a more specific interpretation of the muscle under
investigation.

Average PCr recovery times \taupcr\ in SOL were shorter than in GM.
Acidification slows down PCr recovery in muscle tissue
\cite{Iotti:1993,Kemp:1993b}, therefore the maximum oxidative capacity
\Qmax\ is a more suitable measure to compare recovery from exercise in
muscles with different pH.  This parameter was also comparable between
SOL and GM, but slightly lower in SOL.
Nevertheless, it has to be taken into account that the two muscles
were working in a different metabolic regime, SOL at low activation
and alkaline pH working mainly aerobic, and GM at much higher
activation and acidification with a high glycolytic contribution.
A sound comparison of these muscles would require similar PCr
depletion and end-exercise pH values, e.g., via multiple measurements
at different exercise intensities, which is beyond the scope of
this study.

The reproducibility measurements show good agreement of data acquired
from SOL in two bouts on the same day (see Table~\ref{tab:repro}).
The stability of mean forces indicates that the subjects were
exercising at a comparable level in both bouts, and similar
end-exercise values of PCr depletion and pH show that SOL was also
activated to nearly the same extent.  \taupcr\ showed small variation
from bout to bout in subjects \#6 and \#9, with larger variability in
subject \#11, the differences from bout to bout were not statistically
significant. All recovery times in SOL were in the lower range of all
measurements in this study and hence distinct from values typically
measured in GM.
When testing day-to-day reproducibility over the course of one week,
PCr depletion was consistently very low and pH was elevated throughout
the exercise in SOL on both days and in all exercise bouts, which were
performed with the same force (see Table~\ref{tab:repro}, subject
\#10).  Reproducing the same effect of exercise (i.e., equal
activation of SOL and hence PCr depletion) by repeating the
measurements after a longer period of several months was attempted in
three subjects and found to be challenging.  Possible reasons for
larger variability (data not shown) are that the subjects may have
distributed the exercise load to individual calf muscles differently,
when asked to perform the same exercise after a long time, and that
training state and living circumstances can change over longer periods
\cite{Hoff:2013}, which could influence the measured parameters.  Yet,
all measurements performed to test reproducibility in SOL resulted in
values that corresponded to low metabolic activation, which confirms
that the striking differences observed in data from SOL and GM were
not caused by repeating the exercise within 25\,min, but represent
intrinsic differences in the metabolic response of these muscles to
the same exercise.  This was also consistent with previous studies
\cite{Baligand:2010,Meyerspeer:2012}, which indicate that repeating
exercise has no severe (if any) influence on PCr recovery kinetics,
after a sufficiently long phase of inactivity. This applies in
particularly to SOL which was activated only modestly.

The main advantage of the semi-LASER sequence is that the size and
orientation of the single voxel can be explicitly defined in all three
dimensions. Consequently, a large voxel can be selected, anatomically
matched to the muscle, which minimizes partial volume effects.
Therefore, the method has high specificity \cite{Meyerspeer:2011},
yields good SNR and, as a single-shot sequence, allows high temporal
resolution.
Together, this is greatly beneficial for measurements in a superficial
muscle like GM. Furthermore it forms the basis of an inherent
virtue of the semi-LASER sequence: the ability to penetrate into
deeper lying tissues and afford accurate, localized data. Obviously, in
deeper areas localization via a small coil is not applicable, and the
location and non-trivial geometry of a muscle like SOL make it
particularly inaccessible, even to other localization methods. With a
2D ISIS-like sequence for instance, as used in
\cite{Vandenborne:1993}, placing voxels in SOL without overlap to GM
or GL is nearly impossible.

Quantification of fast pH changes requires high temporal resolution,
e.g., to follow the immediate pH increase at the onset of exercise.
In GM, the average time to maximum pH was 23\,s, and some subjects
showed a similarly fast increase of pH in SOL. Such dynamics can only be
resolved with a temporal resolution of 11\,s or better, to avoid
undersampling. 
Also the quantification of the PCr recovery time constant \taupcr\
greatly benefits from higher temporal resolution.  
Particularly at low depletions, a sufficient number of data points
needs to be acquired while PCr concentration is still significantly
below resting values, i.e., the low dynamic range necessitates high
time resolution.
Aided by the temporal resolution of 6\,s, the feasibility of \taupcr\
quantification at PCr depletions as low as 15\,\% was demonstrated in
this study.

The initial pH rise supports the notion that in the first seconds of
exercise PCr breakdown is the dominant mechanism to ATP production,
with mitochondrial activity not being elevated beyond resting
activity, and minimal glycolytic contribution.
The alkaline pH throughout the exercise in SOL suggests that the ATP
need can be met by PCr breakdown and oxidative phosphorylation
\cite{Kemp:1994}, which is in agreement with the modest PCr depletion
rates. 
A small decrease of post exercise pH below pre exercise levels in SOL
was found within all groups (i.e., with the full exercise protocol and
in reproducibility measurements), which can be explained by the proton
production associated with oxidative ATP synthesis after aerobic
exercise.

The elevated pH in SOL also affirms that the PCr decrease (and Pi
increase) is a physiologic effect and not due to a motion artifact or
contamination from strongly depleted adjacent muscles, since this
would be accompanied by a pH drop or a split Pi peak. 
Contamination is a form of spatial averaging across two muscles, which
can result in artifacts of the apparent Pi peak, regarding position,
width and shape, when the muscles are in different metabolic states.
Consequently the resulting value of pH can lie anywhere between the
actual pH values of the respective muscles, depending on
the volume fraction of the muscles in the averaged volume (see
Fig.~\ref{fig:var-avg}c and d).  This bias of pH can also vary with
time, due to the different metabolic evolution of the muscles during
exercise.
Also temporal averaging can be a source of artifacts, as the Pi peak
moves in frequency, when pH changes over time.  With low temporal
resolution this is perceived as an increased Pi linewidth, or when the
(averaged) acquisition is sufficiently long compared to the pH shift,
even might appear as two components in the broad peak.

To summarize, accurate spatial localization and sufficiently high
temporal resolution help to establish a more reliable, precise and
accurate quantification of Pi and hence pH, specific to a particular
muscle.

\section*{Conclusion}

Localized \xP\ MR spectra were acquired from the deeper lying soleus
muscle at high temporal resolution during and after exercise.
The time courses of PCr and pH indicate lower activation of SOL than
the neighboring GM in this form of exercise.
The high temporal resolution is the basis for quantification of fast
pH changes, e.g., the initial rise of pH which can be on the order of
10\,s.
Furthermore, high temporal resolution is desirable to improve
quantification of PCr recovery, and to avoid erroneous Pi peak
broadening, by minimizing temporal averaging over shifting metabolite
peaks.
All together, this work illustrates the relevance and effect of high
temporal resolution and accurate spatial localization, enabling
quantification of fast or subtle metabolic processes, as well as
improving specificity in many regards.

\vspace{6ex}

\paragraph{Acknowledgments:}\quad\\{\small
This work has been supported by the Austrian BMWFJ, FFG Project
\#832107, ``Vienna Research Studio for Ultra-High Field Magnetic
Resonance Applications'',
Austrian Science Fund (FWF): J\,3031-N20, I\,1743-B13 and an
unrestricted grant to Ewald Moser funded by Siemens Medical.}\\

\paragraph{Conflict of interest:}\quad\\{\small
The authors declare that they have no conflict of interest.}\\

\paragraph{Ethical standards:}\quad\\{\small The study
has been approved by the local ethics committee and has therefore been
performed in accordance with the ethical standards of the
1964 Declaration of Helsinki and its later amendments. All subjects
gave informed consent in writing before being included in the study.

\vspace{6ex} 

\bibliography{bibexport}
\bibliographystyle{magma} 

\end{document}